\documentclass[a4paper]{jpconf}

\usepackage{amssymb}
\usepackage{amsmath}
\usepackage{amscd}
\usepackage{latexsym}

\usepackage{float}

\usepackage{cite}

\usepackage{graphicx}
\usepackage{subfigure}

\newcommand{\be}{\begin{equation}}
\newcommand{\ee}{\end{equation}}

\newcommand{\dlt}{\delta}
\newcommand{\prt}{\partial}
\newcommand{\bfr}{{\bf r}}
\newcommand{\bk}{{\bf k}}
\newcommand{\bfe}{{\bf e}}
\newcommand{\ba}{{\bf a}}
\newcommand{\bp}{{\bf p}}
\newcommand{\bu}{{\bf u}}
\newcommand{\bt}{\beta}
\newcommand{\vp}{\varphi}

\newcommand{\al}{\alpha}

\newcommand{\gm}{\gamma}
\newcommand{\om}{\omega}

\newcommand{\Gm}{\Gamma}
\newcommand{\dgr}{\dagger}
\newcommand{\lbd}{\lambda}

\newcommand{\rgl}{\rangle}
\newcommand{\lgl}{\langle}

\begin{document}

\title{Phonon instability of insulating states in optical lattices}

\author{V.I. Yukalov$^1$ and K. Ziegler$^2$ }

\address{
$^1$Bogolubov Laboratory of Theoretical Physics,
Joint Institute for Nuclear Research, \\ Dubna 141980, Russia}

\address{
$^2$Institut f\"{u}r Physik, Universit\"{a}t Augsburg,
Augsburg D-86135, Germany }

\ead{yukalov@theor.jinr.ru}

\begin{abstract}
The influence of collective phonon excitations, due to intersite
atomic interactions, on the stability of optical lattices is analyzed.
These phonon excitations are shown to essentially reduce the ability
of atoms to be localized. The states that seem to be insulating in
the absence of the phonon excitations can become delocalized when the
latter are present. The delocalization effect exists for both
long-range as well as local atomic interactions.
\end{abstract}

\section{Introduction}

Physics of cold atoms in optical lattices is an intensive field of research,
as can be inferred from the review articles
\cite{Morsh_1,Moseley_2,Bloch_3,Yukalov_4,Yukalov_5}. Depending on the depth
of the wells at the lattice sites, temperature, and interaction strength,
atoms can be in an insulating localized state or in delocalized itinerant
state. Here we consider insulating states.

In an insulating state, an atom, localized at a lattice-site well, being in
the ground state, is described by a well localized atomic wave function, as
in Fig. 1a. This function can exhibit two types of variations. First, the wave
function of the ground state can be transformed into an excited state, e.g.,
as is pictured in Fig. 1b, which corresponds to a single-particle excitation.
Second, because of interactions with other atoms, the wave packet can
oscillate around the lattice site, as is shown in Fig. 1c, which manifests
collective phonon excitations.

Thus, in physical reality, there always exist two types of atomic motion,
the transfer between different energy levels, accompanied by an essential wave
function deformation, and small oscillations around a lattice site, preserving
the wave-function shape.

If one postulates that the system is described by a Hubbard Hamiltonian, this
implies that small oscillations are disregarded and only transitions between
energy bands are taken into account. The Hubbard model is formulated in terms
of single-particle states and does not contain collective atomic fluctuations.
Clearly, there are no phonon excitations in the standard Hubbard model. However,
one should not confuse a model and the physical reality. It is well known that
in any real physical system, whether it is gas, liquid, or solid, there always
exist sound waves caused by particle interactions. And this does not depend
on the presence or absence of any external fields. In nature, any realistic
physical system of interacting atoms does exhibit the presence of density,
or sound, waves characterizing collective phonon excitations. Any model is
a cartoon of reality, and may take or not take into account the existence
of phonons. But one should not confuse cartoon models and physics.

In the present paper, we show how it is possible to modify the Hubbard model
in order to take into account collective phonon excitations. And we analyze
the consequence of the phonon existence on the stability of insulating states.
It turns out \cite{Yukalov_6} that phonon excitations can strongly influence
the region of stability of insulating states in optical lattices. In those
cases, where the phonon instability occurs, the lattice can be metastable,
and living sufficiently long time for being experimentally observed.

\section{Phonon-dressed lattice model}

Let us start with the general form of the Hamiltonian in the second-quantization
representation,
\be
\label{1}
\hat H =
\int \psi^\dgr(\bfr) \left [ -\;
\frac{\nabla^2}{2m} + V_L(\bfr) \right ] \psi(\bfr)\; d\bfr +
\frac{1}{2} \int  \psi^\dgr(\bfr)  \psi^\dgr(\bfr')
\Phi(\bfr-\bfr') \psi(\bfr') \psi(\bfr)\; d\bfr d\bfr ' \; ,
\ee
in which the field operators can satisfy either Bose or Fermi statistics. The
optical-lattice potential
\be
\label{2}
V_L(\bfr) = \sum_{\al=1}^d V_\al\sin^2( k_0^\al r_\al )
\ee
is formed by laser beams with the laser wave vector
\be
\label{3}
\bk_0 = \left \{ k_0^\al = \frac{2\pi}{\lbd_\al} =
\frac{\pi}{a_\al} \right \}
\ee
prescribing the lattice vector
\be
\label{4}
 \ba =  \left \{ a_\al = \frac{\lbd_\al}{2} =
\frac{\pi}{k_0^\al} \right \} \;  .
\ee
The real-space dimensionality is $d = 1,2,3$. The characteristic kinetic
energy of an atom, caused by laser beams, is the recoil energy
\be
\label{5}
  E_R = \frac{k_0^2}{2m} \qquad
\left ( k_0^2 =  \sum_{\al=1}^d \frac{\pi^2}{a_\al^2} \right ) \; .
\ee

The total number of atoms $N$ is placed inside an optical lattice, with
the number of sites $N_L$. The lattice vectors are
$\{ {\bf a}_j: j = 1,2, \ldots, N_L \}$. The lattice itself is fixed in
space. The filling factor
\be
\label{6}
\nu \equiv \frac{N}{N_L} = \rho a^d
\ee
can be expressed through the average atomic density $\rho$ and mean
interatomic distance $a$, defined as
\be
\label{7}
 \rho \equiv \frac{N}{V} \;  , \qquad
a^d \equiv \frac{V}{N_L} \; ,
\ee
with $V$ being the system volume.

Atomic interactions, generally, contain two parts, represented by local
and nonlocal potentials
\be
\label{8}
  \Phi(\bfr) = \Phi_{loc}(\bfr) + \Phi_{non}(\bfr) \; .
\ee
The local interaction potential can be written in the form
\be
\label{9}
 \Phi_{loc}(\bfr) = \Phi_d \dlt(\bfr) \;  ,
\ee
with the interaction strength
\be
\label{10}
 \Phi_d = \frac{\Phi_{eff}}{(\sqrt{2\pi}\; l_\perp)^{3-d}} \; , \qquad
\Phi_{eff} \equiv 4\pi \; \frac{a_{eff}}{m} \;  ,
\ee
in which $l_\perp$ is the transverse oscillator length and $a_{eff}$ is an
effective scattering length. The latter, depending on the system geometry,
takes the following values: in the quasi-one-dimensional case \cite{Olshanii_7},
it is
$$
 a_{eff} = \frac{a_s}{1-0.46 a_s/l_\perp} \qquad ( d = 1 ) \;  .
$$
In the quasi-two-dimensional case \cite{Petrov_8}, one has
$$
a_{eff} =
\frac{a_s}{1- ( a_s/\sqrt{2\pi} \;l_\perp)\ln [(2\pi)^{3/2}\rho l_\perp a_s]}
\qquad ( d = 2 ) \;   .
$$
And in three dimensions, it is just the scattering length,
$$
 a_{eff} = a_s \qquad (d = 3 ) \;  .
$$
Nonlocal interactions can correspond to dipolar forces
\cite{Griesmaier_9,Baranov_10}

In an insulating state, atoms are localized at the points of minima of
an effective potential formed by the combination of the given optical
lattice and an effective potential created by other atoms. At each given
moment of time the points of minima $\{ {\bf r}_j \}$ do not necessarily
coincide with the lattice sites $\{ {\bf a}_j \}$, since atoms fluctuate.
The set $\{ {\bf r}_j \}$ of the effective minima is to be treated as a
random set. As far as atoms are localized in the vicinity of the points
$\{ \bfr_j \}$, it is possible to expand the field operators over
localized orbitals \cite{Harrison_11,Slater_12} centered at the
corresponding spatial points,
\be
\label{11}
  \psi(\bfr) =\sum_{nj} c_{nj} \psi_n(\bfr-\bfr_j ) \; ,
\ee
where $n$ is a set of quantum numbers defining energy levels. Substituting
expansion (\ref{11}) into Hamiltonian (\ref{1}), we consider only the lowest
energy level, assuming that the gap between the energy levels is sufficiently
large, being much larger than an average phonon energy. Then we obtain the
Hamiltonian
\be
\label{12}
\hat H = - \sum_{i\neq j} J(\bfr_{ij} ) c_i^\dgr c_j +
 \sum_j \left ( \frac{\bp_j^2}{2m} + V_L \right ) c_i^\dgr c_j
+ \frac{U}{2} \sum_j c_j^\dgr c_j^\dgr c_j c_j   +
\frac{1}{2} \sum_{i\neq j} U(\bfr_{ij} ) c_i^\dgr c_j^\dgr c_j c_i \; ,
\ee
in which $J({\bf r}_{ij})$ is a hopping term, ${\bf p}_j^2/2m$ is a
kinetic-energy term, $V_L$ is an average of the lattice potential, $U$ is
an on-site interaction parameter, and $U({\bf r}_{ij)}$ describes
interactions between different sites. Here
${\bf r}_{ij} \equiv {\bf r}_i - {\bf r}_j$. The above quantities are the
matrix elements over the localized orbitals, whose calculation can be found
in Refs. \cite{Yukalov_6,Yukalov_13,Yukalov_14}.

Instead of the localized orbitals, it is possible to use the so-called
maximally localized Wannier functions \cite{Silvestrelly_15,Marzari_16}.
These functions have been successfully employed not only for crystals with
ideally periodic lattices, but also for crystals with defects, disordered
networks, amorphous solids, and even liquids \cite{Silvestrelly_15,Marzari_16}.
The definition of the maximally localized Wannier functions, which can be used
for the present consideration, is given in the Appendix.

Since the vector ${\bf r}_j$, by assumption, is a random vector, close to
${\bf a}_j$, it is convenient to introduce the notation
\be
\label{13}
 \bfr_j = \ba_j + \bu_j \;  ,
\ee
where
\be
\label{14}
 \ba_j \equiv \lgl \bfr_j \rgl \; , \qquad
\lgl \bu_j \rgl = 0 \;  .
\ee
Here the angle brackets mean statistical averaging with respect to the
total system Hamiltonian. In a stable equilibrium state, random oscillations
around the sites ${\bf a}_j$ imply zero average deviation
$\langle {\bf u}_j \rangle$. A nonzero value of the latter would signify a
structural phase transition \cite{Ziegler_17}.

For what follows, it is convenient to define the relative-distance vectors
$\ba_{ij} \equiv \ba_i - \ba_j$ and the relative deviations
$\bu_{ij}\equiv\bu_i-\bu_j$. The deviations are assumed to be small, which
makes it possible to resort to the second-order expansions
$$
U(\bfr_{ij} ) \simeq U_{ij} + \sum_\al U_{ij}^\al u_{ij}^\al  \; - \;
\frac{1}{2} \sum_{\al\bt} U_{ij}^{\al\bt} u_{ij}^\al u_{ij}^\bt \; ,
$$
\be
\label{15}
 J(\bfr_{ij} ) \simeq J_{ij} + \sum_\al J_{ij}^\al u_{ij}^\al  \; - \;
\frac{1}{2} \sum_{\al\bt} J_{ij}^{\al\bt} u_{ij}^\al u_{ij}^\bt \;  ,
\ee
in which
$$
U_{ij} \equiv U(\ba_{ij} ) \; , \qquad
U_{ij}^\al \equiv \frac{\prt U_{ij}}{\prt a_{ij}^\al} =
\frac{\prt U_{ij}}{\prt a_i^\al} \; ,
\qquad
U_{ij}^{\al\bt} \equiv -\; \frac{\prt^2 U_{ij}}{\prt a_{ij}^\al \prt a_{ij}^\bt} =
\frac{\prt^2 U_{ij}}{\prt a_i^\al \prt a_j^\bt} \; ,
$$
with similar notations used for the hopping term.

Keeping in mind different physical nature of deviations and atomic operators,
we employ the decoupling
\be
\label{16}
  u_{ij}^\al u_{ij}^\bt c_i^\dgr c_j =
\lgl u_{ij}^\al u_{ij}^\bt \rgl c_i^\dgr c_j +
 u_{ij}^\al u_{ij}^\bt \lgl c_i^\dgr c_j \rgl -
\lgl u_{ij}^\al u_{ij}^\bt \rgl \lgl  c_i^\dgr c_j \rgl \; .
\ee

Phonon operators are introduced by the nonuniform canonical transformation
\cite{Yukalov_14,Yukalov_18}
$$
\bu_j = \vec{\dlt}_j +
\frac{1}{\sqrt{2N} } \sum_{ks} \sqrt{ \frac{\nu}{m\om_{ks} } } \;
\bfe_{ks} \left ( b_{ks} + b_{-ks}^\dgr \right ) e^{i\bk \cdot \ba_j} \; ,
$$
\be
\label{17}
\bp_j = -\;
\frac{i}{\sqrt{2N} } \sum_{ks} \sqrt{ \frac{m\om_{ks} }{\nu} } \;
\bfe_{ks} \left ( b_{ks} - b_{-ks}^\dgr \right ) e^{i\bk \cdot \ba_j}
\ee
differing from the standard transformation by the presence of the vector
$\vec{\delta}_j$ that is required for cancelling in the Hamiltonian the
terms linear in the phonon operators \cite{Yukalov_19}.

The phonon spectrum is defined by the eigenproblem
\be
\label{18}
  \frac{\nu}{m} \sum_{j(\neq i)}
\sum_\bt \Phi_{ij}^{\al\bt} e^{i\bk\cdot\ba_{ij}} e_{ks}^\bt =
\om_{ks}^2 e_{ks}^\al \; ,
\ee
in which ${\bf e}_{ks}$ are the polarization vectors, $s$ being a
polarization index, and the renormalized dynamical matrix is
\be
\label{19}
\Phi_{ij}^{\al\bt} = U_{ij}^{\al\bt} \lgl c_i^\dgr c_j^\dgr c_j c_i \rgl -
2J_{ij}^{\al\bt} \lgl c_i^\dgr c_j \rgl \;   .
\ee
In the case of a cubic lattice, it is convenient to define the effective
dynamical matrix
\be
\label{20}
  D_{ij} \equiv -\; \frac{1}{d} \sum_{\al=1}^d \Phi_{ij}^{\al\al} \qquad
( D_0 \equiv D_{\lgl ij\rgl }  )  \; ,
\ee
denoting the matrix in Eq. (\ref{20}) for the nearest neighbour sites by $D_0$ .

In this way, we obtain the Hamiltonian
\be
\label{21}
  \hat H = E_N + \hat H_{at} + \hat H_{ph} + \hat H_{ind} \; ,
\ee
where $E_N$ is a non-operator quantity, the second term is the renormalized
atomic Hamiltonian, the third term is the phonon Hamiltonian, and the last
term describes effective atomic interactions induced by the phonon existence.
Expressing these terms, we use the correlation functions
$$
 \lgl u_{ij}^\al u_{ij}^\bt \rgl =
 2 ( 1 - \dlt_{ij} )  \lgl u_j^\al u_j^\bt \rgl \;  ,
$$
$$
\lgl u_i^\al u_j^\bt \rgl = \frac{\dlt_{ij}}{2N_L} \sum_{ks}
\frac{e_{ks}^\al e_{ks}^\bt}{m\om_{ks} } \;
\coth \left ( \frac{\om_{ks}}{2T} \right ) \; .
$$
The first term reads as
\be
\label{22}
  E_N = - N K_N \;  ,
\ee
where $K_N$ is the mean kinetic energy per atom
$$
K_N \equiv \frac{1}{N_L} \sum_j \left \lgl \frac{\bp_j^2}{2m} \right \rgl =
\frac{1}{4\nu N} \sum_{ks} \om_{ks}
\coth \left ( \frac{\om_{ks}}{2T} \right ) \;   .
$$
The atomic Hamiltonian becomes
\be
\label{23}
 \hat H_{at} = - \sum_{i\neq j} \widetilde J_{ij} c_i^\dgr c_j +
\frac{U}{2} \sum_j c_j^\dgr c_j^\dgr c_j c_j +
\frac{1}{2} \sum_{i \neq j}  \widetilde U_{ij} c_i^\dgr c_j^\dgr c_j c_i +
K_N \sum_j c_j^\dgr c_j \;  ,
\ee
with the renormalized hopping and interaction terms
$$
\widetilde J_{ij} =
J_{ij} - ( 1 - \dlt_{ij} )
\sum_{\al\bt} J_{ij}^{\al\bt} \lgl u_j^\al u_j^\bt \rgl \; ,
\qquad
 \widetilde U_{ij} =
U_{ij} - ( 1 - \dlt_{ij} )
\sum_{\al\bt} U_{ij}^{\al\bt} \lgl u_j^\al u_j^\bt \rgl \;  .
$$
The phonon Hamiltonian is
\be
\label{24}
 \hat H_{ph} = \sum_{ks} \om_{ks}
\left ( b_{ks}^\dgr b_{ks} + \frac{1}{2} \right ) \;  ,
\ee
in which the phonon spectrum is defined in Eq. (\ref{18}) through the
renormalized dynamical matrix (\ref{19}). And the induced atomic
interactions result in the term
\be
\label{25}
 \hat H_{ind} = \sum_{i\neq j}
 \sum_{\al\bt} F_i^\al \gm_{ij}^{\al\bt} F_j^\bt \;  ,
\ee
where
$$
F_i^\al =\sum_{j(\neq i)} \left ( 2 J_{ij}^\al c_i^\dgr c_j -
U_{ij}^\al c_i^\dgr c_j^\dgr c_j c_i \right ) \; , \qquad
\gm_{ij}^{\al\bt} = \frac{1}{N_L}
\sum_{ks} \frac{e_{ks}^\al e_{ks}^\bt}{m\om_{ks}^2} \; e^{i\bk\cdot\ba_{ij}} \;   .
$$

\section{Lindemann criterion of stability}

An insulating state, by its definition, presupposes that atomic wave
packets are well localized close to their lattice sites, so that the
packets, corresponding to the nearest neighbours, practically do not
overlap. As soon as the overlap becomes essential, the system cannot
anymore be treated as localized. In that case, the insulating state
is not stable, but there occurs the delocalization of atoms. 
This behavior can be discussed in terms of correlation functions:
The correlation of the fluctuations in the localized state should 
decay exponentially on a finite correlation length $\xi$ \cite{ziegler15}. 
Then an instability due to an increasing overlap is indicated by a divergency 
of the correlation length. A similar description is based on the Lindemann 
criterion, which will be used subsequently. 

The oscillations of an atomic wave packet are characterized by the
mean-square deviation
\be
\label{26}
r_0^2 \equiv \sum_{\al=1}^d \lgl u_j^\al u_j^\al \rgl \;   .
\ee
In an insulating state, this deviation is usually much smaller than
the distance between the nearest neighbours $a$. The delocalization
transition happens, when the mean square deviation increases approaching
$a$. This is the Lindemann criterion of delocalization \cite{Lindemann_20}.
The weak form of the criterion implies that
\be
\label{27}
 \frac{r_0}{a} < 1 \;  .
\ee

We calculate the mean-square deviation for a cubic lattice, with
nearest-neighbour interactions, and employ the Debye approximation. The
lattice is assumed to be sufficiently large, so that $N_L \gg 1$. Then,
at finite temperature, we have in one dimension
\be
\label{28}
r_0^2 \simeq \frac{T}{2\pi^2 \nu D_0} \; N_L \qquad ( d = 1, \; T>0 ) \; ,
\ee
in two dimensions
\be
\label{29}
 r_0^2 \simeq \frac{T}{(2\pi)^2 \nu D_0} \; \ln N_L
\qquad ( d = 2, \; T>0 ) \;   ,
\ee
and in three dimensions
\be
\label{30}
 r_0^2 \simeq \frac{9T}{m T_D^2} \; N_L \qquad ( d = 3, \; T>T_D ) \;   ,
\ee
where the Debye temperature is
\be
\label{31}
 T_D = \sqrt{ 4\pi\; \frac{\nu D_0}{m} } \;
\left [ \frac{d}{2} \; \Gm\left ( \frac{d}{2} \right ) \right ]^{1/d} \;  .
\ee
This tells us that, at finite temperature, only three-dimensional insulating
states can be stable.

At zero temperature, we find
\be
\label{32}
  r_0^2 = \frac{d^2}{2(d-1)mT_D}  \qquad (  T = 0 ) \;  ,
\ee
which shows that, at $T = 0$, two-dimensional and three-dimensional
insulating states are admissible.

Summarizing, we see that insulating optical lattices can be stable with
respect to collective phonon excitations: in two dimensions at zero
temperature, if
\be
\label{33}
  \frac{E_R}{T_D} < \frac{\pi^2}{2}  \qquad ( d=2, \; T = 0 ) \;  ,
\ee
in three dimensions at zero temperature when
\be
\label{34}
  \frac{E_R}{T_D} < \frac{2\pi^2}{3}  \qquad ( d=3, \; T = 0 ) \;  ,
\ee
and in three dimensions at finite temperature, provided that
\be
\label{35}
  \frac{E_R T}{T_D^2} < \frac{\pi^2}{6}  \qquad ( d=3, \; T > T_D ) \;  .
\ee
The meaning of these conditions is rather clear. They require that kinetic
energy be smaller than potential energy by the amount in the right-hand
side of these inequalities.

Even if the insulating state in an optical lattice is not stable, it can
be metastable, living quite long time. The lifetime of a metastable state
can be estimated by the formula
$$
t_{met} = \frac{2\pi}{\om_0} \; \exp \left ( \frac{V_0}{E_R} \right ) \; ,
$$
in which $\omega_0$ is the effective oscillator frequency corresponding
to a potential well at a lattice site, $V_0$ is the optical lattice
barrier height, and $E_R$ is the recoil energy. For example, in the case
of a cubic optical lattice filled by $^{87}$Rb atoms \cite{Greiner_21},
the lifetime of the insulating state is quite long, being $t_{met} > 200$ s,
which is longer than the lifetime of atoms in a trap, which allows one to
accomplish the necessary measurements. The lifetime of insulating states
for atoms, such as $^{52}$Cr, $^{162}$Er, and $^{164}$Dy, possessing
long-range dipole interactions \cite{Griesmaier_9,Baranov_10}, can be
even longer.

In conclusion, taking into account collective phonon degrees of freedom can
essentially change the region of stability of insulating states in optical
lattices \cite{Yukalov_6}. The phonon instability can be characterized by
the Lindemann criterion \cite{Lindemann_20}. It looks that the insulating
optical lattices, studied at the present time in experiments with trapped
atoms, are not stable with respect to phonon excitations, but correspond
to only metastable states that, although, can live sufficiently long time
for being experimentally observed.

We have introduced the phonons in complete analogy with their introduction
in the theory of quantum crystals \cite{Guyer_22}, where, first, one
considers atoms, localized in some randomly distributed spatial points
close to the sites of a periodic lattice. Then, defining small deviations
from these lattice sites, one introduces phonon excitations. The sole
difference between the optical lattices and quantum crystals is that the
lattice sites in the former are prescribed by laser beams creating the
lattice, while in the case of quantum crystals, the lattice vectors have
to be defined by minimizing the system free energy.

Note that sound waves exist in all interacting systems of many particles,
whether in ideally periodic crystals or amorphous solids, in normal or
superfluid liquids, in bulk samples or finite systems \cite{Srivastava_25,Birman_23}.

\section*{Appendix: Maximally localized Wannier functions}

The localized Hamiltonian (\ref{12}) has been derived by employing
the field-operator expansion (\ref{11}) over the localized orbitals
describing single-particle states of atoms localized in the vicinity
of the related lattice sites. We have also mentioned that, instead of
the localized orbitals, we could used the maximally localized Wannier
functions \cite{Silvestrelly_15,Marzari_16}. Here we give a brief
definition of these functions in the form that would be convenient
for the purpose of our paper.

Suppose, first, that atoms are localized in a lattice described by
a set of lattice vectors $\{ {\bf r}_j: j=1,2,\ldots, N_L \}$. Because
of atomic interactions, the minima of an effective potential, defining
the lattice $\{ {\bf r}_j \}$, do not coincide with the sites of the
optical lattice $\{ {\bf a}_j \}$, although each ${\bf r}_j$ is close
to ${\bf a}_j$. One can define a Bloch function
$$
 \vp_{nk}(\bfr) = u_{nk}(\bfr) e^{i\bk\cdot\bfr} \; ,
\qquad
u_{nk}(\bfr+\bfr_j) = u_{nk}(\bfr) \; .
$$
However, Bloch functions are strongly nonunique, since the new function
$$
\overline\vp_{nk}(\bfr) = \sum U_{mn}(\bk) \vp_{mk}(\bfr)
$$
is also another Bloch function, provided that the matrix $[U_{mn}]$
is unitary, such that
$$
\sum_j U_{jm}^*(\bk) U_{jn}(\bk) = \sum_j U_{mj}(\bk) U_{nj}^*(\bk)
= \dlt_{mn} .
$$
It is straightforward to check that the new Bloch functions are orthonormal
and form a complete basis,
$$
 \int \overline\vp_{mk}^*(\bfr) \overline\vp_{np}(\bfr) \; d\bfr =
\dlt_{mn} \dlt_{kp} \; , \qquad
\sum_{nk} \overline\vp_{nk}(\bfr) \overline\vp_{nk}^*(\bfr') =
\dlt(\bfr-\bfr') \; .
$$

Respectively, Wannier functions are also strongly nonunique, enabling
the introduction of the form
$$
  \psi_n(\bfr-\bfr_j) \equiv \frac{1}{\sqrt{N_L}} \;
\sum_k \overline\vp_{nk}(\bfr)  e^{-i\bk\cdot\bfr_j}  =
\frac{1}{\sqrt{N_L}}
\sum_{mk} U_{mn}(\bk) \vp_{mk}(\bfr) e^{-i\bk\cdot\bfr_j} =
$$
$$
\frac{1}{\sqrt{N_L}}
\sum_{mk} U_{mn}(\bk) u_{mk}(\bfr) e^{i\bk\cdot(\bfr-\bfr_j)} =
\frac{1}{\sqrt{N_L}}
\sum_{mk} U_{mn}(\bk) u_{mk}(\bfr-\bfr_j) e^{i\bk\cdot(\bfr-\bfr_j)}\;  .
$$
Maximally localized Wannier functions are defined as the functions, with
the matrix $[U_{mn}]$ minimizing the variance functional
$$
\sum_n \left ( \lgl r^2 \rgl_n - \lgl r \rgl_n^2 \right ) \;   ,
$$
where the brackets $<\cdots>_n$ imply the notation
$$
\lgl A(\bfr) \rgl_n \equiv \int \psi_n^*(\bfr) A(\bfr) \psi_n(\bfr)\; d\bfr \;  .
$$
The so-defined maximally localized Wannier functions are orthogonal
to each other, being strongly localized, and exponentially decaying
outside of their related centers ${\bf r}_j$.

These well localized Wannier functions can be used for deriving the localized
model (\ref{11}). Then, taking into account that each location ${\bf r}_j$
is close to the site ${\bf a}_j$, small deviations are introduced as in
Eqs. (\ref{13}) and (\ref{14}). The deviations describe atomic fluctuations
corresponding to collective phonon excitations.

\section*{Acknowledgement}

Financial support from RFBR (grant $\#$14-02-00723) and from the University
of Augsburg is appreciated.

\begin{figure}[ht]
\vspace{9pt}
\centerline{
\hbox{ \includegraphics[width=4.75cm]{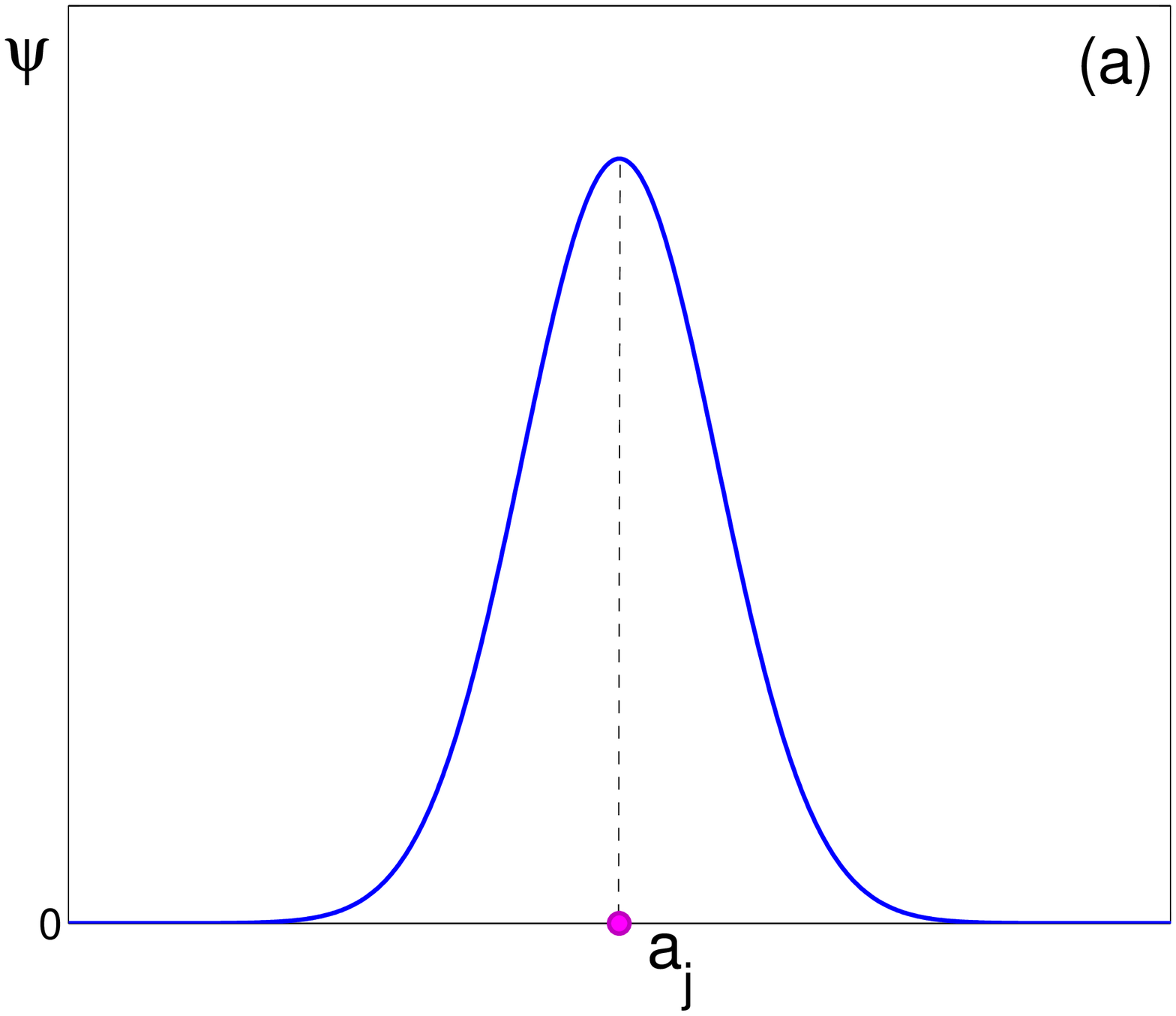} \hspace{0.5cm}
\includegraphics[width=4.75cm]{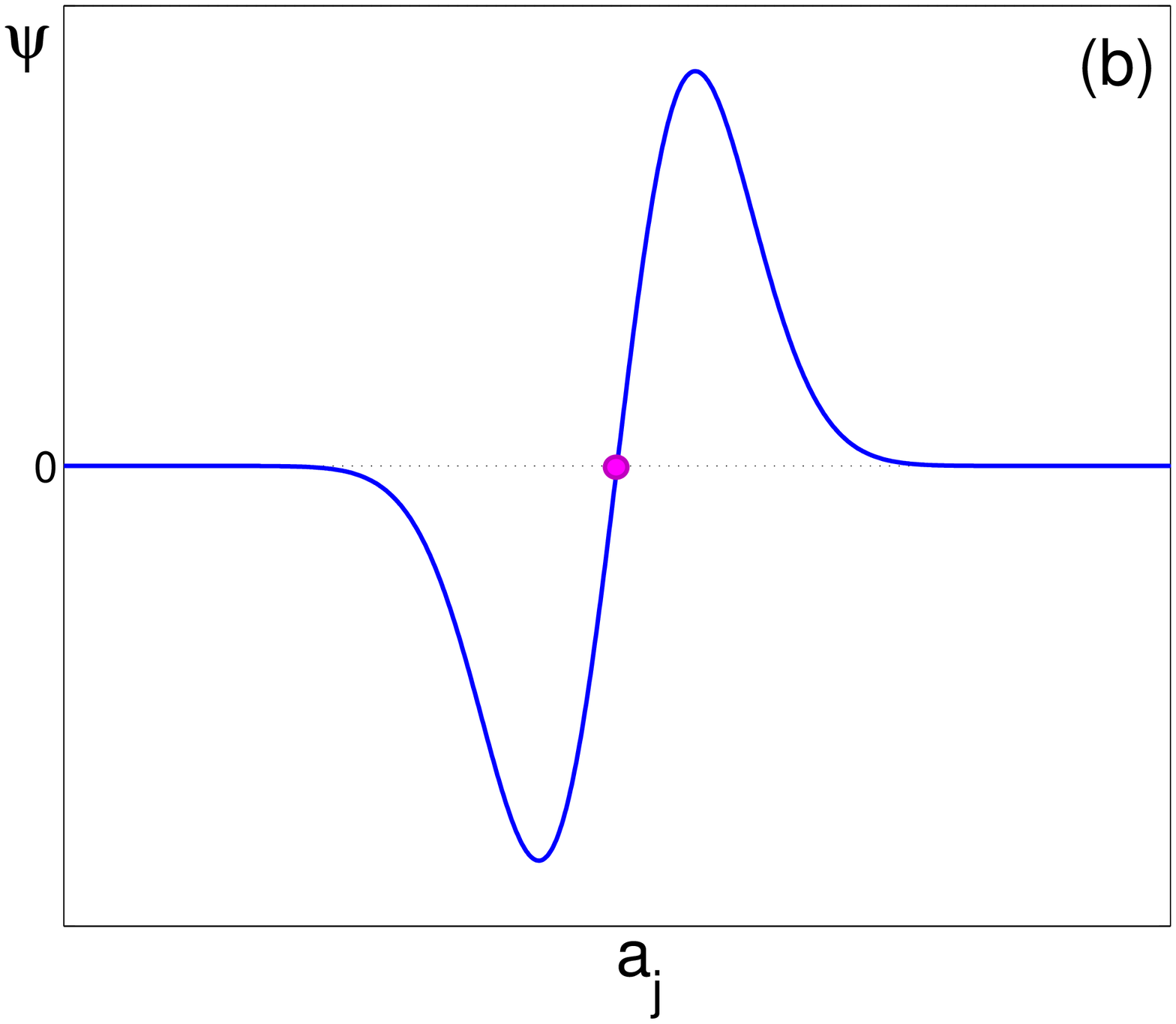}  \hspace{0.5cm}
\includegraphics[width=4.75cm]{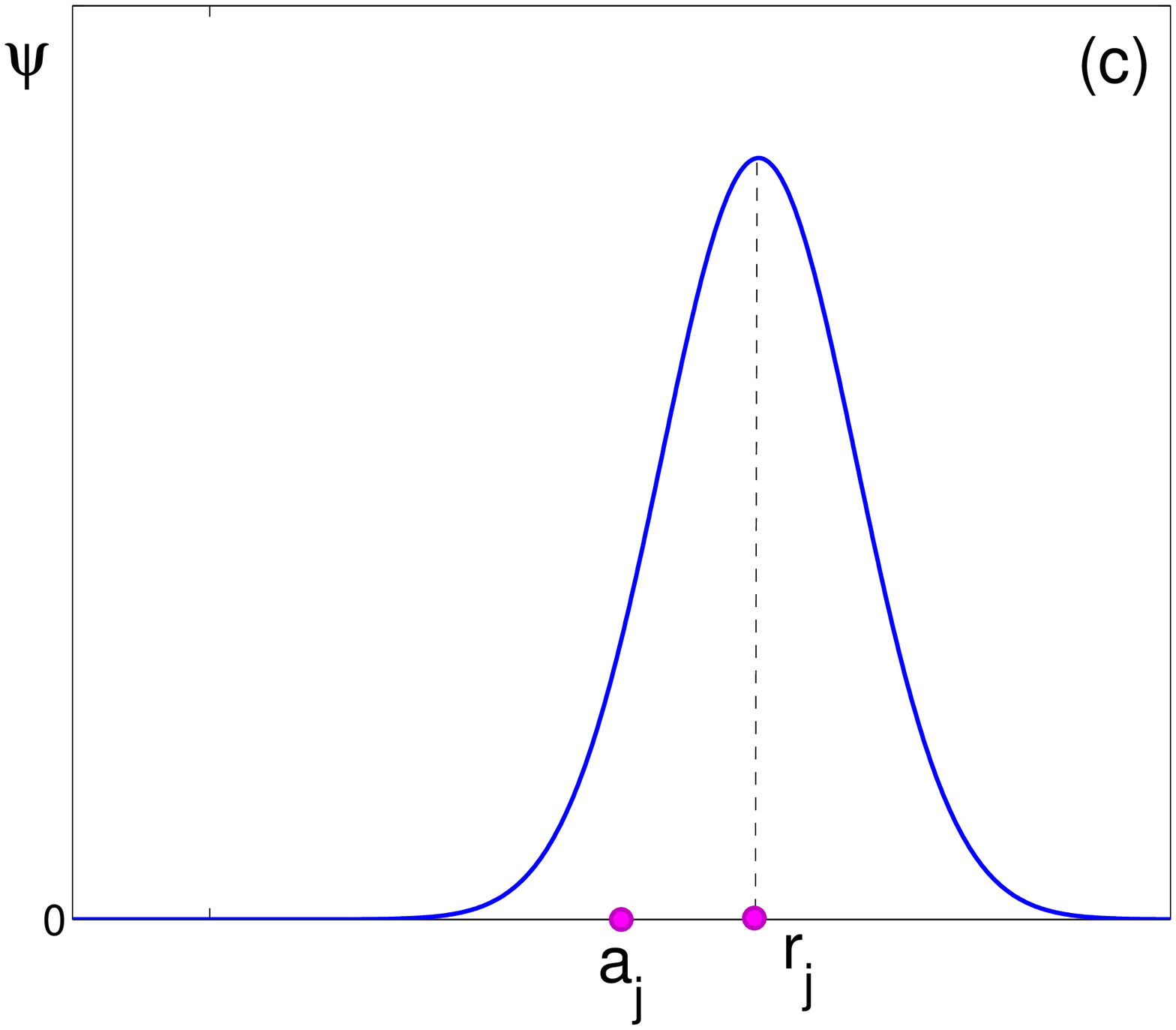} } }
\caption{
(a) The wave function for the ground state of an atom localized at
a lattice site $a_j$;
(b) The wave function for a localized atom in an excited
single-particle state;
(c) The wave function of an atom displaced from the lattice site $a_j$
to a close position $r_j$, such that $|\bfr_j - \ba_j| \ll a$, with
$a$ being the nearest neighbour lattice distance.}
\label{fig:Fig.1}
\end{figure}

\end{document}